\documentclass[twocolumn,showpacs,final,prd]{revtex4}
\usepackage{graphicx}
\newcommand{\be}{\begin{equation}}
\newcommand{\ee}{\end{equation}}

\begin{document}
\title{The construction of a general inner product in non-Hermitian quantum theory and\\some explanation for the nonuniqueness of the ${\cal C}$ operator in ${\cal PT}$ quantum mechanics}
\author{Frieder Kleefeld}
\email{kleefeld@cfif.ist.utl.pt}
\affiliation{Collaborator of the Centro de F\'{\i}sica das Interac\c{c}\~{o}es Fundamentais,
Instituto Superior T\'{e}cnico, Technical University of Lisbon,
P-1049-001 Lisboa Codex, Portugal\\
Present address for postal correspondence: Pfisterstr. 31, 90762 F\"urth, Germany}
\date{\today}

\begin{abstract}
Most recently it has been observed e.g.\ by Bender and Klevansky (arXiv:0905.4673 [hep-th] \cite{Bender:2009en}) that the ${\cal C}$-operator related to a ${\cal PT}$-symmetric non-Hermitian Hamilton operator is not unique. Moreover it has been remarked by Shi and Sun (arXiv:0905.1771 [hep-th] \cite{Shi:2009pc}) very recently that there seems to exist a well defined inner product in the context of the Hamilton operator of the ${\cal PT}$-symmetric  non-Hermitian Lee model yielding a different ${\cal C}$-operator as compared to the one previously derived by Bender {\em et al.}. The puzzling observations of both manuscripts are reconciled and explained in the present manuscript as follows: the actual form of the metric operator (and the induced ${\cal C}$-operator) related to some non-Hermitian Hamilton operator constructed along the lines of Shi and Su depends on the chosen normalization of the left and right eigenvectors of the Hamilton operator under consideration and is therefore ambiguous. For a specific ${\cal PT}$-symmetric $2\times 2$-matrix Hamilton operator it is shown that --- by a suitable choice of the norm of its eigenvectors --- the metric operator yielding a positive semi-definite inner product can be made even independent of the parameters of the considered Hamilton operator. This surprising feature makes in turn the obtained metric operator rather unique and attractive. For later convenience the metric operator for the Bosonic and Fermionic (anti)causal harmonic oscillator is derived.
\end{abstract}

\pacs{03.65.-w,11.30.-j,11.10.Ef,03.65.Ge}

%
\maketitle

\section{Introduction}
The formulation of a consistent quantum-theoretical framework for physical systems described by non-Hermitian Hamiltonians is for various reasons \cite{Kleefeld:2005hf,Kleefeld:2005hd,Kleefeld:2002au,Kleefeld:2004jb,Kleefeld:2004qs,Kleefeld:1999xxyy,Znojil:2001mc,Kleefeld:2006bp,Kleefeld:2003dx,Kleefeld:2003zj,Kleefeld:2002gw,Kleefeld:2001xd,Kleefeld:1998yj,Kleefeld:1998dg} some very topical and important issue which is known to have a long history 
\cite{Kleefeld:2002au,Kleefeld:2004jb,Kleefeld:2004qs,Kleefeld:1999xxyy,Znojil:2001mc,Kleefeld:2006bp,Bender:2007nj,Bender:2005tb}.

Important progress by the application of non-Hermitian Hamilton operators is to be expected in the theory of strong interactions (and the intimately related
theory of superconductivity) where hadronic meson-meson and meson-nucleon scattering is well described \cite{Kleefeld:2003bw,Kleefeld:2005hd,vanBeveren:2005pk,Kleefeld:2002au} over a wide range of energies by some non-Hermitian interaction converting hadrons into intermediate (anti)quarks the confining effective interaction Hamiltonian of which is displaying also strong signals of non-Hermiticity. That such type of non-Hermiticity of a Hamilton operator does not necessarily spoil the unitarity and probability concept of the underlying quantum theory will be clarified to some extent below.

Moreover had it been discoveries in mathematical physics \cite{Caliceti:1980,Fernandez:1998} noticing that several non-Hermitian Hamilton operators are sharing various features with (often non-local or non-linear) Hermitian Hamilton operators making these specific non-Hermitian operators interesting candidates to describe the time evolution and spectral properties of specific physical systems. 

Stimulated by some important conjecture by Bessis (and Zinn-Justin) of 1992 on the reality and positivity of spectra for manifestly non-Hermitian Hamiltonians received the research field considering non-Hermitian Hamilton operators some considerable boost \cite{PTworkshop:2003} when Bender and Boettcher proposed in 1997 \cite{Bender:1998ke} that the reality property of spectra of Hamilton operators is connected with their anti-unitary \cite{robnik1986} ${\cal PT}$-symmetry, i.e.\ symmetry under simultaneous space (${\cal P}$, parity) reflection and time (${\cal T}$) reversal. The reality and boundness of the spectrum for eigenstates with unbroken ${\cal PT}$-symmetry has been proven meanwhile rigorously for a general class of ${\cal PT}$-symmetric Hamilton operators \cite{Dorey:2001uw}. Unfortunately it had become also clear that the ${\cal PT}$ operation cannot be used as a metric for a quantum mechanical inner product as the resulting pseudo-norm would be indefinite \cite{Kleefeld:2004jb,Kleefeld:2004qs,Znojil:2001mc,Japaridze:2001py,Bagchi:2001qu,Weigert:2003py,Kleefeld:2006bp,Mostafazadeh:2002hb,Blasi:2003xxyy,Trinh:2005zr,Tanaka:2006ja,Mostafazadeh:2004mx,Scholtz:1992xxyy,Bender:2005tb,Lee:1969fy}. This noted indefiniteness caused researchers working in the field to recall and review the long history of the concept of pseudo-Hermiticity  and indefinite metrics \cite{Kleefeld:2004jb,Kleefeld:2004qs,Znojil:2001mc} going back to names like P.A.M.\ Dirac, W.\ Pauli, W.\ Heisenberg, L.S.\ Pontrjagin, M.G.\ Kre\u{\i}n, $\ldots$ in order to find a suitable inner product with positive semi-definite norm for functions subject to ${\cal PT}$-symmetric Hamilton operators replacing the  well known inner product with positive semi-definite norm associated with the name Max Born \cite{Pais:1982we} being applicable only in the context of Hermitian Hamilton operators. The identification of a so-called ${\cal CPT}$-inner product with positive semi-definite norm by Bender, Brody and Jones \cite{Bender:2002vv} (see e.g.\ also discussions in Refs.\ \cite{Bender:2004sa,Bender:2004ej,Mostafazadeh:2003az,Samsonov:2005}) led then finally to a well defined pseudo-Hermitian quantum mechanics \cite{Bender:2007nj,Bender:2005tb,Bender:1998gh,Mostafazadeh:2008pw} and quantum field theory \cite{Bender:2004sa} for {\em non-Hermitian ${\cal PT}$-symmetric} Hamilton operators.

As attractive the discovery by Bender {\em et al.} has been that physical systems described by non-Hermitian ${\cal PT}$-symmetric Hamiltonians obtain a quantum theoretic foundation by the construction of some ${\cal C}$-operator yielding a probability concept via an induced ${\cal CPT}$-inner product \cite{Bender:2002vv,Bender:2005tb}, as cumbersome turned out the actual exact or just perturbative construction of some ${\cal C}$-operator even for simple non-Hermitian Hamiltonians \cite{Jones:2006qs,Jones:2004gp,Mostafazadeh:2005wm,Bender:2004by,Bender:2006fg}. In this spirit it has been a great success that 
Bender {\em et al.} \cite{Bender:2004sv,Bender:2007nj} were e.g.\ able to determine exactly some ${\cal C}$-operator for the non-Hermitian ${\cal PT}$-symmetric Lee-model \cite{Lee:1954iq,Lee:1969fz,Kleefeld:2004jb}, while Jones \cite{Jones:2007pq} constructed subsequently from the non-Hermitian ${\cal PT}$-symmetric Hamiltonian of the Lee-model its respective Hermitian counterpart  by a simililarity transform. 

Facing the very successful concept of a ${\cal CPT}$-inner product for physical systems described by non-Hermitian ${\cal PT}$-symmetric Hamiltonians it came to us as a surprise when Shi and Sun suggested in their most recent article ``Recovering Unitarity of Lee Model in Bi-Orthogonal Basis'' \cite{Shi:2009pc} some inner product for the non-Hermitian ${\cal PT}$-symmetric Lee-model seemingly different from the aforementioned ${\cal CPT}$-inner product by Bender {\em et al.} which shares all the attractive features of the ${\cal CPT}$-inner product including a probability concept. 

Moreover turned it out to be very puzzling to us when Bender and Klevansky noted in a very recent article \cite{Bender:2009en} (see also \cite{Mostafazadeh:2005wm}) that the ${\cal C}$-operator of a ${\cal PT}$-symmetric quantum theory is not unique facing the situation that Das and Greenwood \cite{Das:2009kx} were able to reproduce the result of  Bender {\em et al.} \cite{Bender:2004sv} on the Lee-model by different means.

This article is devoted to reconcile the observations of Bender {\em et al.} with the results of Shi and Sun and to try to explain why Bender and Klevansky observed that  the ${\cal C}$-operator of a ${\cal PT}$-symmetric quantum theory is not unique.

The article is organized as follows. In Section \ref{sec2} we recall the definition of the general inner product for non-Hermitian Hamiltonians used by Shi and Sun \cite{Shi:2009pc}. In Section \ref{sec3} we apply the definition of the inner product mentioned in the article by Shi and Su to a system described by some ${\cal PT}$-symmetric $2\times2$-matrix Hamilton operator originally suggested and discussed by Bender {\em et al.}. Section \ref{sec4} is used to revisit the non-Hermitian Hamilton operator for the Bosonic and Fermionic (anti)causal harmonic oscillator and to derive its metric operator along the lines described in Section \ref{sec2}. In Section \ref{secsum1} we summarize our results and conclude.

\section{A general inner product for a non-Hermitian Hamiltonian} \label{sec2}

Let's consider some Hamilton operator $H$ which is not necessarily Hermitian, i.e. we assume particularly the situation \mbox{$H\not=H^+$} with $H^+$ being the Hermitian conjugate of $H$. The right eigenvectors $\left|e_n\right>$ and left eigenvectors $\left<\!\left<e_n\right|\right.$ for the respective (sometimes even complex valued) eigenvalues $E_n$ (The index $n\in\{0,1,2,\ldots\}$ labels the eigenvalue!) respect the following well known equations:
\begin{eqnarray} H \left|e_n\right> & = & E_n \,  \left|e_n\right> \; , \\[2mm]
\left<\!\left<e_n\right|\right.\! H & = & E_n \, \left<\!\left<e_n\right|\; ,\right.
\end{eqnarray}
implying also the following well known identy:
\begin{eqnarray} 0 & = & \left<\!\left<e_{n^\prime}\right|\right.\! H \left|e_n\right> - \left<\!\left<e_{n^\prime}\right|\right.\!  H \left|e_n\right> \nonumber \\[2mm] 
& = & \Big( \left<\!\left<e_{n^\prime}\right|\right.\! H\Big) \left|e_n\right> - \left<\!\left<e_{n^\prime}\right|\right.\! \Big( H \left|e_n\right>\Big) \nonumber \\[2mm] 
 & = & (E_{n^\prime}-E_n) \left<\!\left<e_{n^\prime}\right|e_n\right> \,
\end{eqnarray}
yielding $\left<\!\left<e_{n^\prime}\right|e_n\right>=0$ for  $E_{n^\prime}\not=E_n$. In what follows we define the following Hermitian conjugation of state vectors according to the following identities:
\begin{eqnarray} \left<e_n\right| & \equiv & \Big( \left|e_n\right> \Big)^+  \; , \\[2mm]
\left.\left|e_n\right>\!\right>  & \equiv & \Big(  \left<\!\left<e_n\right|\right.\! \Big)^+  \; ,
\end{eqnarray}
implying of course also $\left<e_{n^\prime}\!\left|e_n\right>\!\right>=\Big(\left<\!\left<e_{n^\prime}\right|e_n\right>\Big)^\ast=0$ for  $E_{n^\prime}\not=E_n$ while $(\ldots)^\ast$ is indicating complex conjugation.

As pointed out in the article by Shi and Sun \cite{Shi:2009pc} one can introduce ``normalized" right eigenvectors  $\left|E_n\right>$ and left eigenvectors $\left<\!\left<E_n\right|\right.$ for the respective eigenvalues $E_n$ of the Hamilton operator $H$ by performing the following definitions (see also \cite{Znojil:2007xxyy}) \footnote{Shi and Sun \cite{Shi:2009pc} use the notation $\left|e_n\right>$, $\left<d_n\right|$, $\left|E_n\right>$, $\left<D_n\right|$ for our respective states $\left|e_n\right>$, 
$\left<\!\left<e_n\right|\right.$, $\left|E_n\right>$, $\left<\!\left<E_n\right|\right.$.}:
\begin{eqnarray} \left|E_n\right> & \equiv & \frac{\left|e_n\right>}{\sqrt{\left<\!\left<e_n\right|e_n\right>}} \; , \label{normev1} \\[2mm] 
\left.\left|E_n\right>\!\right> & \equiv & \frac{\left.\left|e_n\right>\!\right>}{\sqrt{\left<e_n\!\left|e_n\right>\!\right>}}  \; ,  \label{normev2}
\end{eqnarray}
yielding
\begin{eqnarray} \left<E_n\right| & \equiv & \frac{\left<e_n\right|}{\sqrt{\left<e_n\!\left|e_n\right>\!\right>}} \; , \label{normev3} \\[2mm] 
\left<\!\left<E_n\right|\right. & \equiv & \frac{\left<\!\left<e_n\right|\right.}{\sqrt{\left<\!\left<e_n\right|e_n\right>}}  \; , \label{normev4}  
\end{eqnarray}
and
\begin{equation} \left<E_{n^\prime}\!\left|E_n\right>\!\right>=\left<\!\left<E_{n^\prime}\right|E_n\right>= \delta_{n^\prime\!,n}\; , \label{eqorth1} \end{equation}
and of course by definition also:
\begin{eqnarray} H \left|E_n\right> & = & E_n \,  \left|E_n\right> \; , \\[2mm]
\left<\!\left<E_n\right|\right.\! H & = & E_n \, \left<\!\left<E_n\right|\; .\right.
\end{eqnarray}
Shi and Sun \cite{Shi:2009pc} then recall the following well known spectral expansion of the Hamilton operator $H$ (see also \cite{Weigert:2003py}):
\begin{equation} H = \sum\limits_n E_n \, \left|E_n\right>\left<\!\left<E_n\right|\right. \; , \label{eqspec1} \end{equation}
the following completeness relations (see also \cite{Weigert:2003py}):
\begin{equation} 1 = \sum\limits_n \left|E_n\right>\left<\!\left<E_n\right|\right. = \sum\limits_n \left.\left|E_n\right>\!\right> \left<E_n\right| \; , \label{eqcomp1} \end{equation}
and suggest the following metric operator $\eta$:
\begin{equation} \eta \equiv \sum\limits_n \left.\left|E_n\right>\!\right>\left<\!\left<E_n\right|\right. = \eta^+ \; , \label{eqmetric1}\end{equation}
yielding obviously:
\begin{eqnarray} \left.\left|E_n\right>\!\right> & = &  \eta \, \left|E_n\right>  \;, \\[2mm]  
\left<\!\left<E_n\right|\right. & = &  \left<E_n\right| \, \eta^+ =   \left<E_n\right| \, \eta \;.
\end{eqnarray}
We would like to add here that the inverse metric operator $\eta^{-1}$ defined by $\eta \, \eta^{-1}= \eta^{-1}\eta \equiv 1$ is given by:
\begin{equation} \eta^{-1} = \sum\limits_n \left|E_n\right>\left<E_n\right| = \Big(\eta^{-1}\Big)^+ \; . \end{equation}
In our aforementioned notation Shi and Sun \cite{Shi:2009pc} (see also \cite{Znojil:2007xxyy,Znojil:2009xxyy}) write down the following inner product for two arbitrary state vectors $\left|\psi\right>$ and $\left|\varphi\right>$  ($\left< \varphi \right|\equiv\left| \varphi\right>^+$):
\begin{eqnarray} \left< \varphi \right. \! \left| \psi\right>_\eta & \equiv & \left< \varphi \right|  \eta\,  \left| \psi\right> \\[2mm]
 & = &  \sum\limits_n \left< \varphi \left|E_n\right>\!\right>\left<\!\left<E_n\right| \psi\right> \\
 & = & \left<\!\left< \varphi \right. \! \left| \psi\right>\right. = \left. \left< \varphi \right. \! \left| \psi\right>\!\right> \\[2mm]
 & = &  \left<\!\left< \varphi \right| \right.\! \eta^{-1} \!\!\left.\left| \psi\right>\!\right> = \left<\!\left< \varphi \right. \! \left| \psi\right>\!\right>_{\eta^{-1}} \; .
\end{eqnarray}
As in standard quantum mechanics the probability of a normalized state $\left|\psi\right>$ (normalization condition: \mbox{$\left< \psi \right. \! \left| \psi\right>_\eta=1$}) having the energy $E_n$ is given by the modulous square 
$|\!\left<\!\left<E_n\right| \psi\right>\!|^2$ of the probability amplitude 
$\left<\!\left<E_n\right| \psi\right>=(\left< \psi \left|E_n\right>\!\right>)^\ast$ as there holds:
\begin{eqnarray} \left< \psi \right. \! \left| \psi\right>_\eta & \equiv & \left< \psi \right|  \eta\,  \left| \psi\right> = \left<\!\left< \psi \right. \! \left| \psi\right>\right. = \left. \left< \psi \right. \! \left| \psi\right>\!\right> \\[2mm]
 & = &  \sum\limits_n \left< \psi \left|E_n\right>\!\right>\left<\!\left<E_n\right| \psi\right> \\
 & = &  \sum\limits_n \left|\left<\!\left<E_n\right| \psi\right>\right|^2 \; .
\end{eqnarray}

\section{Revisiting a ${\cal PT}$-symmetric matrix Hamiltonian of Bender {\em et al.}} \label{sec3}

To illustrate why the ${\cal C}$-operator should be ambiguous and not unique we consider the following simple $2\times 2$-matrix Hamilton operator suggested and studied thoroughly by Bender, Brody and Jones in Ref.\ \cite{Bender:2002vv} (see also \cite{Bender:2004xxyy,Bender:2005tb,Geyer:2007xxyy}):
\begin{equation} H = \left( 
\begin{array}{cc} r \,e^{i\,\theta} & s \\[2mm] 
t & r \,e^{-i\,\theta} 
\end{array}
\right) \end{equation}
with $r$, $s$, $t$ and $\theta$ being real parameters. As has been pointed out in Ref.\ \cite{Bender:2002vv} the Hamilton operator is obviously ${\cal PT}$-symmetric, as there holds the following identity:
\begin{equation} H^+ = {\cal P} \, H \, {\cal P} \; ,
\end{equation}
with ${\cal P}$ being the respective parity operator defined as follows:
\begin{equation} {\cal P} = \left( 
\begin{array}{cc} 0 & 1 \\[2mm] 
1 & 0 
\end{array}
\right) \; . \label{parop1}
\end{equation}
The Hamilton operator $H$ has two eigenvalues $E_+$ and $E_-$ which have been determined already in Ref.\ \cite{Bender:2002vv} and which are given by:
\begin{equation} E_\pm  =  r\,\cos \theta \pm \sqrt{st - r^2\,\sin^2 \theta} \; .
\end{equation}
As has been pointed also out already in Ref.\ \cite{Bender:2002vv} the eigenvalues $E_\pm$ of $H$ are real provided that there holds:
\begin{equation} st \ge r^2\,\sin^2 \theta \, . 
\end{equation}
The right eigenvectors $\left|e_\pm\right>$ and left eigenvectors $\left<\!\left<e_\pm\right|\right.$ corresponding to the eigenvalues $E_\pm$ are --- up to some normalization --- determined by the following conditions:
\begin{eqnarray} H \left|e_\pm\right> & = & E_\pm \,  \left|e_\pm\right> \; , \\[2mm]
\left<\!\left<e_\pm\right|\right.\! H & = & E_\pm \, \left<\!\left<e_\pm\right|\; ,\right.
\end{eqnarray}
yielding with the definition $\sin\alpha \equiv \frac{r}{\sqrt{st}}\,\sin\theta$ of Ref.\ \cite{Bender:2002vv} 
\begin{eqnarray} \left|e_\pm\right> & \propto & 
\left( 
\begin{array}{r}  \sqrt{s} \; e^{\pm i \frac{\alpha}{2} } 
\\[2mm]  \pm \sqrt{t} \; e^{\mp i \frac{\alpha}{2}}  
\end{array}
\right) \; , \label{eqprop1} \\[2mm]
\left<\!\left<e_\pm\right|\right. & \propto & \Big( \sqrt{t} \; e^{\pm i \frac{\alpha}{2} } , \pm \sqrt{s} \; e^{\mp i \frac{\alpha}{2}} \Big) \; . \label{eqprop2}
\end{eqnarray}
The crucial point for the nonuniqueness of the ${\cal C}$-operator and therefore also the metric operator $\eta$ is to our understanding related to the fact that {\em normalization of the right eigenvectors $\left|e_\pm\right>$ and left eigenvectors $\left<\!\left<e_\pm\right|\right.$ of $H$ is not unique}, yet can be chosen conveniently. To explain this point we will choose without loss of generality now two different suitable normalizations for the aforementioned eigenvectors to obtain respectively two suitable versions of the ${\cal C}$-operator:\\[3mm]
{\em Case 1: The ${\cal C}$-operator chosen by Bender et al.}\\[3mm]
Throughout this case we will assume the proportionality constants in Equations (\ref{eqprop1}) and (\ref{eqprop2}) to be unity and choose the right eigenvectors $\left|e_\pm\right>$ and left eigenvectors $\left<\!\left<e_\pm\right|\right.$ of $H$ to be therefore given exactly by the following expressions:
\begin{eqnarray} \left|e_\pm\right>  & = &  
\left( 
\begin{array}{r}  \sqrt{s} \; e^{\pm i \frac{\alpha}{2} } 
\\[2mm]  \pm \sqrt{t} \; e^{\mp i \frac{\alpha}{2}}  
\end{array}
\right) \; , \\[2mm]
\left<\!\left<e_\pm\right|\right. & = & \Big( \sqrt{t} \; e^{\pm i \frac{\alpha}{2} } , \pm \sqrt{s} \; e^{\mp i \frac{\alpha}{2}} \Big) \; ,
\end{eqnarray}
 yielding by Hermitian conjugation of course
\begin{eqnarray}
\left<e_\pm\right| & = & \Big( \sqrt{s} \; e^{\mp i \frac{\alpha}{2} } , \pm \sqrt{t} \; e^{\pm i \frac{\alpha}{2}} \Big) \; , \\[2mm]
 \left.\left|e_\pm\right>\!\right> & = & 
\left( 
\begin{array}{r}  \sqrt{t} \; e^{\mp i \frac{\alpha}{2} } 
\\[2mm]  \pm \sqrt{s} \; e^{\pm i \frac{\alpha}{2}}  
\end{array}
\right) \; , 
\end{eqnarray}
and consequently also
\begin{equation} \left<\!\left<e_\pm\right|e_\pm\right> = \left<e_\pm\left|e_\pm\right>\!\right> = 2\,\sqrt{st}\;\cos\alpha \; .
\end{equation}
Using these results and Equations (\ref{normev1}),  (\ref{normev2}), (\ref{normev3}) and (\ref{normev4}) we obtain for the respective normalized right eigenvectors $\left|E_\pm\right>$ and left eigenvectors $\left<\!\left<E_\pm\right|\right.$ of $H$ the following expressions:
\begin{eqnarray} \left|E_\pm\right>  & = & \frac{1}{\sqrt{2\,\sqrt{st}\;\cos\alpha}} \,  
\left( 
\begin{array}{r}  \sqrt{s} \; e^{\pm i \frac{\alpha}{2} } 
\\[2mm]  \pm \sqrt{t} \; e^{\mp i \frac{\alpha}{2}}  
\end{array}
\right) \; , \\[2mm]
\left.\left|E_\pm\right>\!\right> & = & \frac{1}{\sqrt{2\,\sqrt{st}\;\cos\alpha}} \,
\left( 
\begin{array}{r}  \sqrt{t} \; e^{\mp i \frac{\alpha}{2} } 
\\[2mm]  \pm \sqrt{s} \; e^{\pm i \frac{\alpha}{2}}  
\end{array}
\right) \; , \\[2mm]
\left<E_\pm\right| & = & \frac{1}{\sqrt{2\,\sqrt{st}\;\cos\alpha}} \,\Big( \sqrt{s} \; e^{\mp i \frac{\alpha}{2} } , \pm \sqrt{t} \; e^{\pm i \frac{\alpha}{2}} \Big) \; , \\[2mm]
\left<\!\left<E_\pm\right|\right. & = & \frac{1}{\sqrt{2\,\sqrt{st}\;\cos\alpha}} \,\Big( \sqrt{t} \; e^{\pm i \frac{\alpha}{2} } , \pm \sqrt{s} \; e^{\mp i \frac{\alpha}{2}} \Big) \; .
\end{eqnarray}
Using these expressions it is straight forward to verify the orthonormality conditions Equation (\ref{eqorth1}), the completeness relations Equation (\ref{eqcomp1}) and the spectral expansion of the Hamilton operator Equation (\ref{eqspec1}). For latter convenience the metric operator $\eta$ defined by Equation (\ref{eqmetric1}) will be now derived explicitely:

\begin{eqnarray} \eta & = &  \left.\left|E_+\right>\!\right> \! \left<\!\left<E_+\right|\right. + \left.\left|E_-\right>\!\right> \!\left<\!\left<E_-\right|\right. \nonumber \\[2mm] 
 & = & \frac{1}{2\,\sqrt{st}\;\cos\alpha} \times \nonumber \\[2mm] 
 & & \Bigg[ \left( 
\begin{array}{r}  \sqrt{t} \; e^{- i \frac{\alpha}{2} } 
\\[2mm]  + \sqrt{s} \; e^{+ i \frac{\alpha}{2}}  
\end{array}
\right) 
\Big( \sqrt{t} \; e^{+ i \frac{\alpha}{2} } , + \sqrt{s} \; e^{- i \frac{\alpha}{2}} \Big)  \nonumber \\[2mm]
 & & + \left( 
\begin{array}{r}  \sqrt{t} \; e^{+ i \frac{\alpha}{2} } 
\\[2mm]  - \sqrt{s} \; e^{- i \frac{\alpha}{2}}  
\end{array}
\right) 
\Big( \sqrt{t} \; e^{- i \frac{\alpha}{2} } , - \sqrt{s} \; e^{+ i \frac{\alpha}{2}} \Big) \Bigg] \nonumber \\[2mm] 
 & = & \frac{1}{2\,\sqrt{st}\;\cos\alpha} \;  \Bigg[ \left( 
\begin{array}{cc} t & \sqrt{st}\,e^{- i \alpha} \\ [2mm] 
\sqrt{st}\,e^{+ i \alpha} & s \end{array}
\right) \nonumber \\[2mm] 
 & & \makebox[1.8cm]{} + \left( 
\begin{array}{cc} t & -\sqrt{st}\,e^{+ i \alpha} \\ [2mm] 
-\sqrt{st}\,e^{- i \alpha} & s \end{array}
\right) \Bigg] \nonumber \\[2mm] 
 & = & \frac{1}{\cos\alpha}   \left( 
\begin{array}{ccc} \sqrt{t/s} & & -i\,\sin\alpha\\ [2mm] 
i\,\sin\alpha & & \sqrt{s/t} \end{array}
\right)
\; . \label{eqmetric2}
\end{eqnarray}
The ${\cal C}$-operator derived by Bender {\em et al.} in Ref.\ \cite{Bender:2002vv} is now obtained as the product of the parity operator Equation (\ref{parop1}) and the metric operator:
\begin{equation} {\cal C} = {\cal P}\, \eta = \frac{1}{\cos\alpha}   \left( 
\begin{array}{ccc} i\,\sin\alpha & & \sqrt{s/t} \\ [2mm] 
\sqrt{t/s} & & -i\,\sin\alpha \end{array}
\right)
\; ,
\end{equation}
provided one assumes $s=t$ yielding $\sqrt{s/t}=\sqrt{t/s}=1$. The disadvantage of this ${\cal C}$-operator and the related metric operator $\eta$ in Equation (\ref{eqmetric2}) is that they depend --- contrary to the metric operator of the inner product of standard quantum mechanics --- on parameters $r$, $s$, $t$ and $\theta$ of the Hamilton operator, i.e.\ they depend on specific properties of the physical system described by the Hamilton operator. The question arises at this point whether it is possible to construct some metric operator $\eta$ and some related ${\cal C}$-operator which are independent of the parameters of the Hamilton operator $H$. The answer to this question will be given during the study of the following case 2.\\[3mm]
{\em Case 2: Parameter independent choice of the ${\cal C}$-operator}\\[3mm]
Throughout this case we will assume the proportionality constants in Equations (\ref{eqprop1}) and (\ref{eqprop2}) to be unity for right eigenvectors and $\pm 1$ for left eigenvectors of $H$. We choose therefore the right eigenvectors $\left|e_\pm\right>$ and left eigenvectors $\left<\!\left<e_\pm\right|\right.$ of $H$ to be given exactly by the following expressions:
\begin{eqnarray} \left|e_\pm\right>  & = &  
\left( 
\begin{array}{r}  \sqrt{s} \; e^{\pm i \frac{\alpha}{2} } 
\\[2mm]  \pm \sqrt{t} \; e^{\mp i \frac{\alpha}{2}}  
\end{array}
\right) \; , \\[2mm]
\left<\!\left<e_\pm\right|\right. & = & \pm \Big( \sqrt{t} \; e^{\pm i \frac{\alpha}{2} } , \pm \sqrt{s} \; e^{\mp i \frac{\alpha}{2}} \Big) \nonumber \\[2mm]
 & = & \Big( \pm \sqrt{t} \; e^{\pm i \frac{\alpha}{2} } ,  \sqrt{s} \; e^{\mp i \frac{\alpha}{2}} \Big) \; ,
\end{eqnarray}
 yielding by Hermitian conjugation of course
\begin{eqnarray}
\left<e_\pm\right| & = & \Big( \sqrt{s} \; e^{\mp i \frac{\alpha}{2} } , \pm \sqrt{t} \; e^{\pm i \frac{\alpha}{2}} \Big) \; , \\[2mm]
 \left.\left|e_\pm\right>\!\right> & = & 
\left( 
\begin{array}{r}  \pm \, \sqrt{t} \; e^{\mp i \frac{\alpha}{2} } 
\\[2mm]   \sqrt{s} \; e^{\pm i \frac{\alpha}{2}}  
\end{array}
\right) \; , 
\end{eqnarray}
and consequently also
\begin{equation} \left<\!\left<e_\pm\right|e_\pm\right> = \left<e_\pm\left|e_\pm\right>\!\right> = \pm \,2\,\sqrt{st}\;\cos\alpha \; . \label{eqinnprod1}
\end{equation}
The interesting aspect of this case is that the left eigenvector of $H$ for the eigenvalue $E_-$ contains an extra factor $(-1)$ as compared to the one of case 1 yielding now the leading minus sign in the inner product on the right-hand side of Equation (\ref{eqinnprod1}). This extra minus sign leads to the fact that one of the eigenvalues of the resulting ${\cal C}$-operator will have now an opposite sign as compared to the eigenvalues of the ${\cal C}$-operator of case 1.
Hence, the original task of the ${\cal C}$-operator of case 1 to render the inner product $\left<\psi|\psi\right>_\eta$ positive semi-definite is taken over in case 2 by the norm of specific left eigenvectors of the Hamilton operator $H$.
 
Using now the previous results derived for the case 2 and Equations (\ref{normev1}),  (\ref{normev2}), (\ref{normev3}) and (\ref{normev4}) we obtain for the respective normalized right eigenvectors $\left|E_\pm\right>$ and left eigenvectors $\left<\!\left<E_\pm\right|\right.$ of $H$ the following expressions:
\begin{eqnarray} \left|E_\pm\right>  & = & \frac{1}{\sqrt{\pm\,2\,\sqrt{st}\;\cos\alpha}} \,  
\left( 
\begin{array}{r}  \sqrt{s} \; e^{\pm i \frac{\alpha}{2} } 
\\[2mm]  \pm \sqrt{t} \; e^{\mp i \frac{\alpha}{2}}  
\end{array}
\right) \; , \\[2mm]
\left.\left|E_\pm\right>\!\right> & = & \frac{1}{\sqrt{\pm\,2\,\sqrt{st}\;\cos\alpha}} \,
\left( 
\begin{array}{r} \pm \sqrt{t} \; e^{\mp i \frac{\alpha}{2} } 
\\[2mm]  \sqrt{s} \; e^{\pm i \frac{\alpha}{2}}  
\end{array}
\right) \; , \\[2mm]
\left<E_\pm\right| & = & \frac{1}{\sqrt{\pm\,2\,\sqrt{st}\;\cos\alpha}} \,\Big( \sqrt{s} \; e^{\mp i \frac{\alpha}{2} } , \pm \sqrt{t} \; e^{\pm i \frac{\alpha}{2}} \Big)  , \; \\[2mm]
\left<\!\left<E_\pm\right|\right. & = & \frac{1}{\sqrt{\pm\,2\,\sqrt{st}\;\cos\alpha}} \,\Big( \pm \sqrt{t} \; e^{\pm i \frac{\alpha}{2} } , \sqrt{s} \; e^{\mp i \frac{\alpha}{2}} \Big)  .\;\;
\end{eqnarray}
Invoking these expressions it is again straight forward to verify the orthonormality conditions Equation (\ref{eqorth1}), the completeness relations Equation (\ref{eqcomp1}) and the spectral expansion of the Hamilton operator Equation (\ref{eqspec1}). As in case 1 we will derive now also for case 2 the metric operator $\eta$ defined by Equation (\ref{eqmetric1}) explicitely:

\begin{eqnarray} \eta & = &  \left.\left|E_+\right>\!\right> \! \left<\!\left<E_+\right|\right. + \left.\left|E_-\right>\!\right> \!\left<\!\left<E_-\right|\right. \nonumber \\[2mm] 
 & = & \frac{1}{2\,\sqrt{st}\;\cos\alpha} \times \nonumber \\[2mm] 
 & & \Bigg[ \left( 
\begin{array}{r} + \sqrt{t} \; e^{- i \frac{\alpha}{2} } 
\\[2mm]   \sqrt{s} \; e^{+ i \frac{\alpha}{2}}  
\end{array}
\right) 
\Big( +\sqrt{t} \; e^{+ i \frac{\alpha}{2} } ,  \sqrt{s} \; e^{- i \frac{\alpha}{2}} \Big)  \nonumber \\[2mm]
 & & - \left( 
\begin{array}{r}  -\sqrt{t} \; e^{+ i \frac{\alpha}{2} } 
\\[2mm] 
 \sqrt{s} \; e^{- i \frac{\alpha}{2}}  
\end{array}
\right) 
\Big( -\sqrt{t} \; e^{- i \frac{\alpha}{2} } ,  \sqrt{s} \; e^{+ i \frac{\alpha}{2}} \Big) \Bigg] \nonumber \\[2mm] 
 & = & \frac{1}{2\,\sqrt{st}\;\cos\alpha} \;  \Bigg[ \left( 
\begin{array}{cc} t & \sqrt{st}\,e^{- i \alpha} \\ [2mm] 
\sqrt{st}\,e^{+ i \alpha} & s \end{array}
\right) \nonumber \\[2mm] 
 & & \makebox[1.8cm]{} - \left( 
\begin{array}{cc} t & -\sqrt{st}\,e^{+ i \alpha} \\ [2mm] 
-\sqrt{st}\,e^{- i \alpha} & s \end{array}
\right) \Bigg] \nonumber \\ 
 & = &  \left( 
\begin{array}{cc} 0 & 1\\ [2mm] 
1 & 0 \end{array}
\right)
\; , \label{eqmetric3}
\end{eqnarray}
being identical to the parity operator of Equation (\ref{parop1}). As in case 1 the ${\cal C}$-operator of case 2 is obtained as the product of the parity operator Equation (\ref{parop1}) and the metric operator:
\begin{equation} {\cal C} = {\cal P}\, \eta = \left( 
\begin{array}{cc} 1 & 0\\ [2mm] 
0 & 1 \end{array}
\right)
\; .
\end{equation}
The remarkable feature of the matrix representation of the metric operator $\eta$ of Equation (\ref{eqmetric3}) and the resulting trivial ${\cal C}$-operator in case 2 is that both are now --- contrary to case 1 --- {\em independent} of the parameters $r$, $s$, $t$ and $\theta$  of the Hamilton operator, i.e. independent of the properties of the physical system described by the respective Hamilton operator. Moreover is it remarkable that the inner product $\left<\psi|\psi\right>_\eta$ remains by construction positive semi-definite with respect to the {\em parameter dependent} left and right eigenvector basis of the Hamilton operator chosen despite the fact that a scalar product based on the parity operator as the metric operator can be indefinite \cite{Kleefeld:2004jb,Kleefeld:2004qs,Znojil:2001mc,Japaridze:2001py,Bagchi:2001qu,Kleefeld:2006bp,Mostafazadeh:2002hb,Weigert:2003py,Blasi:2003xxyy,Trinh:2005zr,Bender:2005tb,Tanaka:2006ja,Scholtz:1992xxyy} if the basis vectors of the left eigenspace of the Hamiltonian are not chosen properly. Since the eigenvalues of the ${\cal C}$-operator of case 2 are all one the ${\cal C}$-operator of case 2 cannot be any interpreted (as in case 1) to be the intrinsic parity operator as suggested by Bender {\em et al.} e.g.\ in Ref.\ \cite{Bender:2004ej}. 

\section{Revisiting the (anti)causal harmonic oscillator} \label{sec4}

E.g.\ in Refs.\ \cite{Kleefeld:2002au,Kleefeld:2004jb,Kleefeld:2004qs,Kleefeld:2003dx,Kleefeld:2003zj,Kleefeld:2002gw,Kleefeld:2001xd,Kleefeld:1998yj,Kleefeld:1998dg,Kleefeld:1999xxyy} and references therein it has been pointed out by us that the causal evolution of some physical system finding its manifestation e.g.\ in the $i\varepsilon$-prescription in denominators of retarded Green's functions of a causal quantum theory implies some at least infinitesimal non-Hermiticity of the causal Hamilton operator $H_C$ responsible for the causal evolution of the system. In order to restore the overall Hermiticity of the Hamilton operator $H=H_C+H_A=(H)^+$ we added to $H_C$ the anticausal Hamilton operator $H_A=(H_C)^+$ which would describe the anticausal, i.e.\ advanced evolution of some physical system backwards in time and which commutes for consistency reasons with $H_C$, i.e.\ $[H_C,H_A]=0$. Moreover have we noted in Refs.\ \cite{Kleefeld:2002au,Kleefeld:2004jb,Kleefeld:2004qs,Kleefeld:2003dx,Kleefeld:2003zj,Kleefeld:2002gw,Kleefeld:2001xd,Kleefeld:1998yj,Kleefeld:1998dg,Kleefeld:1999xxyy} and references therein that the Hamilton operators describing relativistic Bosons and Fermions, the mass of which is assumed to be complex due to (anti)causal boundary conditions, display the features of the Hamilton operator of some harmonic oscillator with some {\em complex-valued} oscillator frequency $\omega$ with the particular property that creation and annihilation operators of causal excitations (or anticausal excitations) are not Hermitian conjugate to each other. Below we will recall first the definition of the Hamilton operator of the Bosonic and Fermionic (anti)causal harmonic oscillator. Then we will try to construct with the method described in Section \ref{sec2} the metric operator $\eta$ for both cases. The derived metric operator and the related positive semi-definite inner product will find in future convenient applications in the context of some causal or anticausal quantum theory for Bosons and Fermions. It is amusing to note by inspection of the results presented below that the structure of the inner product recalled by Shi and Sun and presented again in Section \ref{sec2} has been anticipated throughout the formulation of the results provided in Refs.\ \cite{Kleefeld:2002au,Kleefeld:2004jb,Kleefeld:2004qs,Kleefeld:2003dx,Kleefeld:2003zj,Kleefeld:2002gw,Kleefeld:2001xd,Kleefeld:1998yj,Kleefeld:1998dg,Kleefeld:1999xxyy}.
\subsection{The Bosonic (anti)causal harmonic oscillator}
The Hamilton operator of the Bosonic (anti)causal harmonic oscillator in 1-dimensional quantum mechanics is given by (see also \cite{Nakanishi:wx,Nakanishi:1972pt,Kleefeld:1999xxyy,Kleefeld:2001xd}):
\begin{equation} H  =  H_C + H_A  =  \frac{1}{2} \, \omega \, \{c^+, a\, \} + \frac{1}{2} \, \omega^\ast  \{a^+, c\, \}  \; . \end{equation}
The relevant Bosonic commutation relations for the causal creation operator ($c^+$), causal annihilation operator ($a$), anticausal creation operator ($a^+$) and anticausal annihilation operator ($c$) are:
\begin{eqnarray} \left( \begin{array}{ll} {[c,c^+]} & {[c,a^+]} \\
{[a,c^+]} & {[a,a^+]} \end{array}\right) & = & \left( \begin{array}{cc} 0 & 1 \\
1 & 0 \end{array}\right)  \; ,  \\[2mm]
 \left( \begin{array}{ll} {[c,c]} & {[c,a]} \\
{[a,c]} & {[a,a]} \end{array}\right) & = & \left( \begin{array}{cc} 0 & 0 \\
0 & 0 \end{array}\right) \; , \\[2mm]
\left( \begin{array}{ll} {[c^+,c^+]} & {[c^+,a^+]} \\
{[a^+,c^+]} & {[a^+,a^+]} \end{array}\right) & = & \left( \begin{array}{cc} 0 & 0 \\
0 & 0 \end{array}\right) \; .
\end{eqnarray}
After defining the conveniently normalized right vacuum state $\left|0\right>$ and left vacuum state $\left<\!\left<0\right|\right.$ by the properties $c\left|0\right>=a\left|0\right>=0$, $\left<\!\left<0\right|\right.c^+=\left<\!\left<0\right|\right.a^+=0$ and $\left<\!\left<0\right|0\right>=1$ the (conveniently normalized) {\em normal} right eigenstates $\left|n,m\right>$ and left eigenstates $\left<\!\left<n,m\right|\right.$ ($n,m\in{\rm I\!N}{}_0$) of $H$ are given by: 
\begin{eqnarray} \left|n,m\right> & = & \frac{1}{\sqrt{n!\,m!}} \,\, (c^+)^n (a^+)^m \left|0\right> \; ,\\[2mm]
\left<\!\left<n,m\right|\right. & = & \frac{1}{\sqrt{m!\,n!}} \, \left<\!\left<0\right|\right. c^m \, a^n \; . 
\end{eqnarray}
They fulfil the following two (stationary) Schr\"odinger equations:
\begin{equation} (H-E_{\,n,m}) \left|n,m\right> = 0, \; \left<\!\left< n,m\right|\right. (H-E_{\,n,m}) = 0 \; ,
\end{equation} 
for the energy eigenvalues: 
\begin{equation} E_{n,m} = \omega  \left(n + \frac{1}{2}\right) + \omega^\ast \left(m + \frac{1}{2}\right)\; .
\end{equation} 
For later convenience we define --- via Hermitian conjugation --- the following adjoint state vectors:
\begin{eqnarray} \left.\left|n,m\right>\!\right> & \equiv & \left<\!\left<n,m\right|\right.^+ = \frac{1}{\sqrt{n!\,m!}} \, (a^+)^n (c^+)^m \!\left.\left|0\right>\!\right>  ,\\[2mm]
\left<n,m\right| & \equiv & \left|n,m\right>^+ = \frac{1}{\sqrt{m!\,n!}} \, \left<0\right| a^m \, c^n \, , 
\end{eqnarray}
and
\begin{equation}\!\left.\left|0\right>\!\right> \equiv \left<\!\left<0\right|\right.^+, \; \left<0\right| \equiv \left|0\right>^+ \; .
\end{equation}

\subsection{The Fermionic (anti)causal harmonic oscillator}
The Hamilton operator of the Fermionic (anti)causal harmonic oscillator in 1-dimensional quantum mechanics is given by (see also \cite{Lee:iw,Kleefeld:1999xxyy,Kleefeld:1998yj,Kleefeld:1998dg}):
\begin{equation} H  =  H_C + H_A  =  \frac{1}{2} \, \omega \, [d^+, b\, ] + \frac{1}{2} \, \omega^\ast  [b^+, d\, ]  \; . \end{equation}
The relevant Bosonic anticommutation relations for the causal creation operator ($d^+$), causal annihilation operator ($b$), anticausal creation operator ($b^+$) and anticausal annihilation operator ($d$) are:
\begin{eqnarray} \left( \begin{array}{ll} {\{d,d^+\}} & {\{d,b^+\}} \\
{\{b,d^+\}} & {\{b,b^+\}} \end{array}\right) & = & \left( \begin{array}{cc} 0 & 1 \\
1 & 0 \end{array}\right) \; , \\[2mm]
 \left( \begin{array}{ll} {\{d,d\}} & {\{d,b\}} \\
{\{b,d\}} & {\{b,b\}} \end{array}\right) & = & \left( \begin{array}{cc} 0 & 0 \\
0 & 0 \end{array}\right) \; ,  \\[2mm]
\left( \begin{array}{ll} {\{d^+,d^+\}} & {\{d^+,b^+\}} \\
{\{b^+,d^+\}} & {\{b^+,b^+\}} \end{array}\right) & = & \left( \begin{array}{cc} 0 & 0 \\
0 & 0 \end{array}\right) \; . 
\end{eqnarray}
After defining the conveniently normalized right vacuum state $\left|0\right>$ and left vacuum state $\left<\!\left<0\right|\right.$ by the properties $d\left|0\right>=b\left|0\right>=0$, $\left<\!\left<0\right|\right.d^+=\left<\!\left<0\right|\right.b^+=0$ and $\left<\!\left<0\right|0\right>=1$ the (conveniently normalized) {\em normal} right eigenstates $\left|n,m\right>$ and left eigenstates $\left<\!\left<n,m\right|\right.$ ($n,m\in \{0,1\}$) of $H$ are given by: 
\begin{eqnarray} \left|n,m\right> & = & \frac{1}{\sqrt{n!\,m!}} \,\, (d^+)^n (b^+)^m \left|0\right> \; ,\\[2mm]
\left<\!\left<n,m\right|\right. & = & \frac{1}{\sqrt{m!\,n!}} \, \left<\!\left<0\right|\right. d^m \, b^n \; . 
\end{eqnarray}
They fulfil the following two (stationary) Schr\"odinger equations:
\begin{equation} (H-E_{\,n,m}) \left|n,m\right> = 0, \; \left<\!\left< n,m\right|\right. (H-E_{\,n,m}) = 0 \; ,
\end{equation} 
for the energy eigenvalues: 
\begin{equation} E_{n,m} = \omega  \left(n - \frac{1}{2}\right) + \omega^\ast \left(m - \frac{1}{2}\right)\; .
\end{equation} 
For later convenience we define --- via Hermitian conjugation --- the following adjoint state vectors:
\begin{eqnarray} \left.\left|n,m\right>\!\right> & \equiv & \left<\!\left<n,m\right|\right.^+ = \frac{1}{\sqrt{n!\,m!}} \, (b^+)^n (d^+)^m \!\left.\left|0\right>\!\right>  ,\\[2mm]
\left<n,m\right| & \equiv & \left|n,m\right>^+ = \frac{1}{\sqrt{m!\,n!}} \, \left<0\right| b^m \, d^n \, , 
\end{eqnarray}
and
\begin{equation}\!\left.\left|0\right>\!\right> \equiv \left<\!\left<0\right|\right.^+, \; \left<0\right| \equiv \left|0\right>^+ \; .
\end{equation}

\subsection{Orthonormality, completeness, metric operator and probability concept}

The following discussion applies to both, the Bosonic {\em and} the Fermionic (anti)causal harmonic oscillator:

The energy eigenvalues $E_{n,n}$ are obviously {\em real}, while the eigenvalues $E_{m,n}$ and $E_{n,m}$ form a {\em complex conjugate} pair for $n\not=m$ which arises typically for the case of broken ${\cal PT}$-symmetry.
In the spirit of Section \ref{sec2} are the {\em (bi)orthogonal} (here {\em normal}) eigenstates complete:
\begin{eqnarray} \left<\!\left< n^\prime,m^\prime\right|n,m\right> & = & \delta_{n^\prime n} \, \delta_{m^\prime m} \; , \\[2mm] 
\sum\limits_{n,m} \left|n,m\right>\!\left<\!\left<n,m\right|\right. &= & 1 \; .
\end{eqnarray}
Following the stategy discussed in Section \ref{sec2} we define the metric operator $\eta$ and the inverse metric operator $\eta^{-1}$ as follows:
\begin{eqnarray} \eta & \equiv & \sum\limits_{n,m} \left.\left|n,m\right>\!\right>\!\left<\!\left<n,m\right|\right. = \eta^+ \; , \\[2mm]
 \eta^{-1} & \equiv & \sum\limits_{n,m} \left|n,m\right>\!\left<n,m\right| = (\eta^{-1})^+ \; .
\end{eqnarray}
As in Section \ref{sec2} the positive semi-definite inner product for two arbitrary state vectors $\left|\psi\right>$ and $\left|\varphi\right>$ ($\left< \varphi \right|\equiv\left| \varphi\right>^+$) is given by:
\begin{eqnarray} \left< \varphi \right. \! \left| \psi\right>_\eta & \equiv & \left< \varphi \right|  \eta\,  \left| \psi\right> \\[2mm]
 & = &  \sum\limits_{n,m} \left< \varphi \left|n,m\right>\!\right>\left<\!\left<n,m\right| \psi\right> \\
 & = & \left<\!\left< \varphi \right. \! \left| \psi\right>\right. = \left. \left< \varphi \right. \! \left| \psi\right>\!\right> \\[2mm]
 & = &  \left<\!\left< \varphi \right| \right.\! \eta^{-1} \!\!\left.\left| \psi\right>\!\right> = \left<\!\left< \varphi \right. \! \left| \psi\right>\!\right>_{\eta^{-1}} \; ,
\end{eqnarray}
while the quantity $|\!\left<\!\left<n,m\right| \psi\right>\!|^2$ can be interpreted to be the probability of a physical system described by the normalized state vector $\left|\psi\right>$ (normalization condition: \mbox{$\left< \psi \right. \! \left| \psi\right>_\eta=1$}) to have the energy $E_{n,m}$. 

\section{Summary and conclusions} \label{secsum1}
As discussed throughout this manuscript and has been recalled in a very recent manuscript by Shi and Sun  \cite{Shi:2009pc} (see also \cite{Znojil:2007xxyy,Znojil:2009xxyy,Znojil:2006xxyy}) there exists a general inner product even for non-Hermitian non-${\cal PT}$-symmetric Hamilton operators which can be defined along the lines described in Section \ref{sec2} of this manuscript by constructing some bi-orthogonal basis of left and right eigenvectors of some non-Hermitian Hamilton operator under consideration. The aforementioned inner product allows to establish some probability concept in the space spanned by the states on which the non-Hermitian Hamilton operator is acting. The very existence of this inner product is --- besides the determination of some complete set of observables --- known to be a crucial ingredient for the construction of some quantum theory. As shown in the present manuscript the construction of the metric operator (and the related ${\cal C}$-operator for some ${\cal PT}$-symmetric Hamilton operator) is not unique, as the normalization of the left and right eigenvectors of the considered non-Hermitian Hamilton operator is not unique. Keeping this in mind one might be tempted to think that there is an infinity of possibilities of establishing a metric operator for a given non-Hermitian Hamilton operator due to the fact that one has an infinity of possibilities to choose the normalization of the left and right eigenvectors of the respective Hamilton operator   (see also \cite{Znojil:2007xxyy,Znojil:2009xxyy,Mostafazadeh:2006xxyy}). Yet the discussion of this manuscript shows also that there exist some metric operators with very particular features which make these metric operators seemingly unique among all other feasible metric operators. The metric operator as originally suggested by Bender {\em et al.}  (see e.g.\ the discussion in Ref.\ \cite{Bender:2004ej}) has the interpretation of some intrinsic parity operator as it assumes the full responsability of rendering the inner product positive semi-definite. Unfortunately does this metric operator explicitely depend on parameters of the Hamilton operator and has to be constructed for each physical system individually (which is often very cumbersome). Fortunately, as demonstrated in the context of case 2 in Section \ref{sec3} of this manuscript, there seems to exist the possibility to normalize the left and right eigenvectors of every non-Hermitian such that the metric operator can be made independent of the parameters of the Hamilton operator. 

A particular --- maybe even unique --- situation like in case 2 of Section \ref{sec3} arises, whenever the choice of the normalization of the left and right eigenvectors of the Hamilton operator yields a trivial ${\cal C}$-operator. A related metric operator $\eta$ will yield the most preferable inner product to set up some quantum theory for a physical system, as it will carry no specific information about the physical system itself (with the exception of some general features of the physical system like e.g.\ the dimension of the phase space involved). {\em The strategy to construct this metric operator suggested by the present manuscript points to the construction and suitable normalization of left and right eigenvectors of some Hamilton operator under consideration.} The question about the existence of such a metric operator is, however, not easy to answer. 

In this context we should recall that the nonuniqueness of the  ${\cal C}$-operator observed by Bender and Klevansky \cite{Bender:2009en} was partially related to the question of how to order operators. It is well known that Green's functions calculated from path integrals or Moyal products \cite{Scholtz:2006xxyy} bearing information of the underlying inner product of some quantum theory \cite{Jones:2009br,Jones:2006sj,Mostafazadeh:2007zza} correspond to a specific kind operator ordering \cite{Testa:1971rb,Rutenberg:1994dn,Dowker:1976xe,Sato:1976hy}
 needed in calculating the same Green's functions on the basis of operator based vacuum-persistence amplitudes. Simultaneously a correct operator ordering is needed to obtain results being Lorentz invariant and displaying no superficial divergencies which cannot be renormalized \cite{Suzuki:1978xz,Dunne:1989gp}. In this spirit the search for a most simple ${\cal C}$-operator/metric operator and the related operator ordering will be guided by the search for Lorentz invariance and renormalizability of the quantum theory under consideration \cite{Dunne:1989gp}. Unfortunately we know that on the level of the action of a quantum theory there can remain some terms called anomalies \cite{Dunne:1989gp,Bender:2006wt,Jones:2006et} which do not find some classical counterpart, the relevance of which in physical laws has to be decided by experiments and the existence of which in a quantum theoretical formalism is related to the specific operator ordering to be used. It is clear that the ambiguity in the ${\cal C}$-operator or metric operator stemming from eventual anomalies should remain persistent in an inner product of any quantum theory and should be visible throughout the determination of some metric operator independent of how simple it finally is chosen by Nature.

Rooted in the work of Scholtz, Geyer and Hahne \cite{Scholtz:1992xxyy} it is nowadays well known that for each ${\cal PT}$-symmetric Hamilton operator $H$  there exists --- with the exception of some very pathologic situations \cite{Bender:2008gh} requiring further investigation --- an infinity \cite{Bender:2008uu} of Hermitian Hamilton operators $h$ related to the former operator $H$ by some similarity transform $h=U  \sqrt{{\cal P}} \, H \sqrt{{\cal P}} \, U^+=h^+$ with $U$ being some arbitrary unitary operator. The ambiguous square root $\sqrt{\cal P}$ of the parity operator ${\cal P}$  Equation (\ref{parop1}) for our Hamilton operator under consideration in Section \ref{sec3} of this manuscript is most conveniently determined from a diagonalized representation of ${\cal P}$ like e.g.\ the following:
\begin{eqnarray} {\cal P} & = & 
 \frac{1}{\sqrt{2}}  
\left( \begin{array}{cc} 1 & 1 \\[2mm] 
1 & -1 
\end{array}
\right)  
\left( 
\begin{array}{cc} 1 & 0 \\[2mm] 
0 & -1 
\end{array}
\right)  \! \frac{1}{\sqrt{2}}  
\left( \begin{array}{cc} 1 & 1 \\[2mm] 
1 & -1 
\end{array}
\right)  , 
\end{eqnarray}
suggesting e.g.
\begin{eqnarray} \sqrt{{\cal P}} & = & 
 \frac{1}{\sqrt{2}}  
\left( \begin{array}{cc} 1 & 1 \\[2mm] 
1 & -1 
\end{array}
\right)  
\left( 
\begin{array}{cc} 1 & 0 \\[2mm] 
0 & \pm \,i 
\end{array}
\right)  \! \frac{1}{\sqrt{2}}  
\left( \begin{array}{cc} 1 & 1 \\[2mm] 
1 & -1 
\end{array}
\right) ,  
\end{eqnarray}
yielding obviously $\sqrt{{\cal P}}^{\,-1} = \sqrt{{\cal P}}^{\,+}$. The very existence of this similarity transform is not at all affected by the nonuniqueness of the ${\cal C}$-operator despite the fact that the actual form of the unitary matrix $U$ can be related to the specific representation of the ${\cal C}$-operator if one establishes the relation ${\cal C}=e^{\cal Q}{\cal P}$ as suggested by Bender {\em et al.} \cite{Bender:2004by,Bender:2004ej,Bender:2003wy,Bender:2005tb}.

For later convenience we have constructed in Section \ref{sec4} along the lines described in Section \ref{sec2} the metric operator and the related positive semi-definite inner product for the Bosonic and Fermionic (anti)causal harmonic oscillators. The results easily generalized to respective Hamilton operators describing (anti)causal Klein-Gordon and Dirac fields will have some crucial importance in formulating a consistent causal, analytical, local and unitary quantum theory for physical systems governed by quasi-Hermitian and non-Hermitian Hamiltonians.

This work was supported by the {\it Funda\c{c}\~{a}o para a
Ci\^{e}ncia e a Tecnologia} \/of the {\it Minist\'{e}rio da Ci\^{e}ncia,
Tecnologia e Ensino Superior} \/of Portugal, under contract
POCI/FP/81913/2007 and by the Czech project LC06002.

\end{document}